%% file: paper.tex
\documentstyle[12pt,epsfig]{article}

\def\KLMUMU{$\mathrm K_L^0 \rightarrow \mu^+ \mu^-$}

\def\KLEE{$\mathrm K_L^0 \rightarrow e^+ e^-$}
\def\KLPINUNU{$\mathrm K_L^0 \rightarrow \pi^0 \nu \overline{\nu}$}
\def\KPPINUNU{$\mathrm K^+ \rightarrow \pi^+ \nu \overline{\nu}$}

\input LP99macros.tex

\def\Title#1{\begin{center} {\Large {\bf #1} } \end{center}}

\begin{document}

\Title{
Quark Mixing Matrix Studies and Lepton Flavor Violation Searches Using
Rare Decays of Kaons}

\bigskip\bigskip


\begin{raggedright}  

{\it William Molzon\index{Molzon, W.}\\ Department of Physics and
Astronomy\\ University of California\\ Irvine, CA 92697-4575  USA}
\bigskip\bigskip
\end{raggedright}

\section{Introduction}

Despite the successes of the Standard Model (SM) of particle physics,
a fundamental understanding of the {\it family structure} of the
quarks and leptons remains elusive. Studies of the kaon system have
long contributed to the limited understanding we do have and experiments continue to
have an impact on our empirical knowledge, despite
the maturity of this  field. In this paper I review
recent results on studies of rare  decays of kaons that contribute in
this area, and discuss future prospects in the field.

Studies of rare decays are set in the context of other recent results
relevant to understanding the family structure of matter. In the quark
sector, two kaon experiments now give consistent results for the ratio
of the level of {\it direct} CP  violation ($\Delta$S = 1 transitions
of $\mathrm K^0_L \rightarrow \pi^+ \pi^-$ and  $\mathrm \pi^0\pi^0)$
to {\it indirect} CP violation ($\Delta$ S = 2 transitions in
$\mathrm K^0 - \overline{K^0}$ mixing),  quantified in the value of
$\mathrm \epsilon^{\prime}/\epsilon$.  The measured ratio is larger
than most SM predictions, but a definitive test of the model
is complicated by hadronic physics uncertainties in the
SM prediction. It is unclear if the measured value can result from a
phase in the quark mixing (CKM) matrix as the SM predicts, and much 
theoretical effort is devoted to answering this important question.

In the lepton sector, compelling evidence for  neutrino mass and
mixing has been found, the strongest being the results on atmospheric
neutrinos, pointing to very large mixing between $\mathrm \nu_{\mu}$
and some other  neutrino, probably $\mathrm \nu_{\tau}$. Neutrino
oscillations also appear to be the only consistent explanation for the low 
flux of solar neutrinos 
observed by a number of experiments. A tremendous experimental effort
is beginning with the goal of measuring fully the lepton mass mixing matrix.

The study of rare decays of kaons potentially contributes to both
these areas.  Measurements of $\mathrm K \rightarrow \pi {\mathit l
\overline{l}}$ decay rates would allow a precise determination of
the parameter that quantifies CP  violation in the SM and an unambiguous test of
the consistency of the SM explanation of CP violation. Measurements 
in the B sector would be compared to those in the K  sector to further
test the SM. In the leptonic  sector, experiments to detect $\mathrm K
\rightarrow (\pi) \mu e$ provide a means to search for lepton flavor
violation (LFV) due to new physics processes other 
than neutrino  mass and mixing.

Measurements of rare decay rates, in particular the small observed rate of
effective flavor changing neutral current (FCNC) decays  and the
absence of LFV decays, have been important in  restricting new physics
models and that  continues to be true. The theoretical motivation for
the experiments has grown  recently with the realization that unified
supersymmetric models can be  constructed that have observable effects
in both these areas and that can be tested in  new experiments. 

In the remainder of this paper I will discuss recent experimental progress
in LFV searches and CKM matrix studies. 

\section{Lepton Flavor Violation}

Experiments to search directly for LFV in the charged sector 
have been performed for many years, all with 
null results. Stringent limits have resulted from searches for
$\mathrm K_L^0 \rightarrow \mu^{\pm} e^{\mp}$~\cite{Arisaka:1993.70,Akagi:1991},
$\mathrm K_L^0 \rightarrow \pi^0 \mu^{\pm}  e^{\mp}$~\cite{Arisaka:1998},
$\mathrm K^+ \rightarrow \pi^+ \mu^+  e^-$~\cite{Lee:1990}, 
$\mathrm \mu^+ \rightarrow e^+ \gamma$~\cite{Bolton:1988,Brooks:1999}, 
$\mathrm \mu^+ \rightarrow e^+e^+e^-$~\cite{Bellgardt:1988}, 
and $\mathrm \mu^- N \rightarrow e^- N$~\cite{Dohmen:1993}. 
The sensitivity of these processes to mechanisms 
that allow LFV varies. In general, kaon decay experiments are most sensitive 
for models that relate lepton and quark family numbers; examples are models 
with leptoquarks, which carry both quark and lepton number.

Two experiments have reported new limits on LFV processes in the kaon
sector: final results of a search for $\mathrm K_L^0 \rightarrow \mu^{\pm} e^{\mp}$
from BNL E871~\cite{Ambrose2:1998} and preliminary results of a 
search for $\mathrm K^+ \rightarrow \pi^+ \mu^+ e^-$ from BNL E865~\cite{Zeller:1999}. 
They are the culmination of rare K decay programs that began about 15 years ago, and no 
experiments that would improve on their sensitivities are currently proposed or 
likely to be proposed soon.

The two modes are related, and differ in the Lorentz structure 
of the underlying physics; the $\mathrm K_L^0$ mode is pseudo-scalar or 
axial-vector, while the $\mathrm K^+$ mode is scalar or vector. 
For a $\mathrm V \pm A$ interaction,
the $\mathrm K^+ \rightarrow \pi^+ \mu^+ e^-$ branching fraction would be smaller by 
a factor of more than 10 by virtue of the larger $\mathrm K^+$ total decay rate and 
the phase space suppression of the 3-body final state. The mass scale to which this
type of experiment is sensitive is easily determined, for example by comparing the 
rate for $\mathrm K^+ \rightarrow \mu^+ \nu_{\mu}$ 
to that for $\mathrm K^0 \rightarrow \mu^+ e^-$ 
(with the exchange of a hypothetical X boson of mass $\mathrm M_X$, coupling g and 
mixing factors $\mathrm \lambda_{sd}$ and $\mathrm \lambda_{\mu e}$).
Forming the ratio of these rates results in the relationship\\ 
\vspace{-.1in}

  \centerline{ $\mathrm 
            M_X ~\simeq ~200~TeV/c^2$ 
           {\large{$~\times$}}
          {\large{ $\mathrm \frac{g_W}{g} ~\times$}}
           {\large{$\mathrm \sqrt{\lambda_{sd} \lambda_{\mu e}} ~\times$}}
            {\large {$\mathrm ( \frac { 10^{-12} }
                     { B ( K_L^0   \rightarrow \mu^{\pm}   e^{\mp}    ) } )^{1/4} $}}  }
\vspace{.05in}

\noindent For example, if $\mathrm B(K_L^0 \rightarrow \mu^{\pm} e^{\mp}) = 10^{-12}$, 
$\mathrm M_X = 200~GeV/c^2$ and $\mathrm g = g_W$, one would infer that  
$\mathrm \lambda_{sd} \lambda_{\mu e} \simeq 10^{-6}$. 

The experimental difficulties of a search with sensitivity $\mathrm 10^{-12}$  
consist of producing enough kaons, building an apparatus with sufficient acceptance 
and rate handling capability, and rejecting processes that could mimic a signal. 
Parameters of a beam line and experiment that do this are a beam with $\mathrm 10^8$ 
$\mathrm K_L^0$ per second, a region with 8\% decay probability, a detector with 
1.5\% acceptance and 3000 hours data collection. The performance is relatively 
insensitive to the kaon beam energy, with the exception that low energy 
($\mathrm < 10$ GeV) allows the use of threshold Cerenkov counters in particle 
identification. Because well defined, clean beams can be made, backgrounds 
result primarily from other K decays, and excellent kinematic measurement and 
particle identification are required to reject them.
Coincidences of two kaon decays, each giving a lepton, are 
also a potential source of background and the detectors must
run at rates up to 1 MHz per detector element and provide good timing signals
to reject them. 

The E871 search for $\mathrm K_L^0 \rightarrow \mu^{\pm} e^{\mp}$ 
was done at  Brookhaven National Laboratory (BNL);
the beam-line and apparatus are described in the literature~\cite{Ambrose2:1998,
Ambrose1:1998, Belz:1999}.  E871 used  a $\mathrm K_L^0$ beam 
produced with a 24 GeV proton beam from the Alternating 
Gradient Synchrotron (AGS). The AGS delivered $\mathrm \sim 1.5 \times 10^{13}$ protons
per 1.5 s spill each 3.6 s, producing a 65 $\mu$str beam of $2 \times 10^8$ K$_L^0$
per pulse in the momentum interval $\mathrm 2 < p_K < 8$ GeV/c. Photons in the 
beam were attenuated using lead foils in a sweeping magnet immediately following 
the target. The beam was produced at an angle of 3.75$^{\circ}$ to minimize the 
neutron flux; the n/K ratio was about 10. This is the most intense neutral kaon 
beam ever made. 

\begin{figure}[b]
\begin{center}
\hbox{\hspace{-.2in} \epsfig{figure=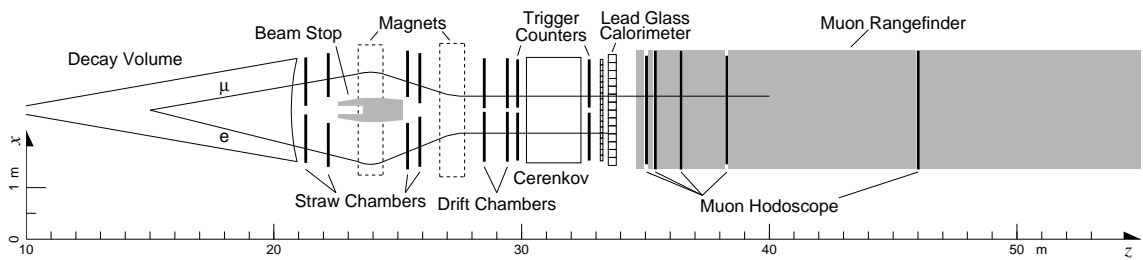,width=6.in}}
\caption{Plan view of the BNL E871 beam line and apparatus.}
\label{e871apparatus}
\end{center}
\end{figure}

Kaons decayed in an 11 m long evacuated tank and the decay products were detected in a 
magnetic spectrometer and system of particle identification counters. 
The apparatus is shown in Figure~\ref{e871apparatus}; it
 differed from that of earlier experiments in the use of a 
beam-stop~\cite{Belz:1999} in the first of two 
analyzing magnets. This reduced rates in downstream detectors and allowed them
to be essentially continuous across the projected beam-line. Important features of the 
spectrometer were the use of
two sequential analyzing magnets to measure redundantly the momenta of charged
decay particles and the use of small diameter straw chambers where rates were highest. 
Redundant means of identifying both electrons and muons were used. 

\begin{figure}[b!]
\begin{center}
\hbox{\epsfig{figure=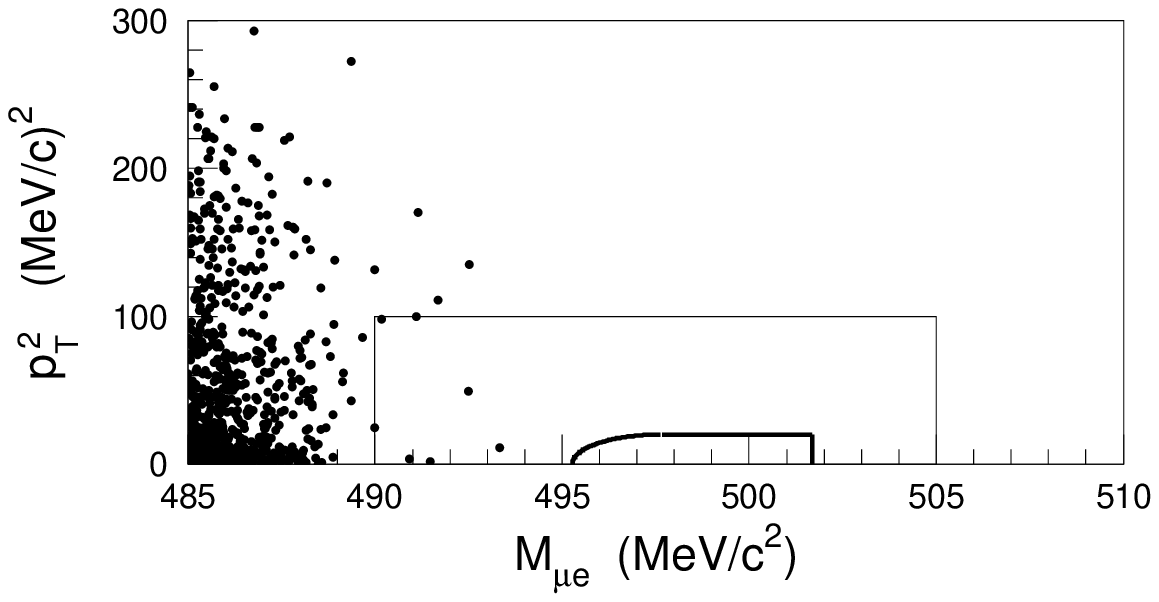,height=1.5in,clip=} 
\epsfig{figure=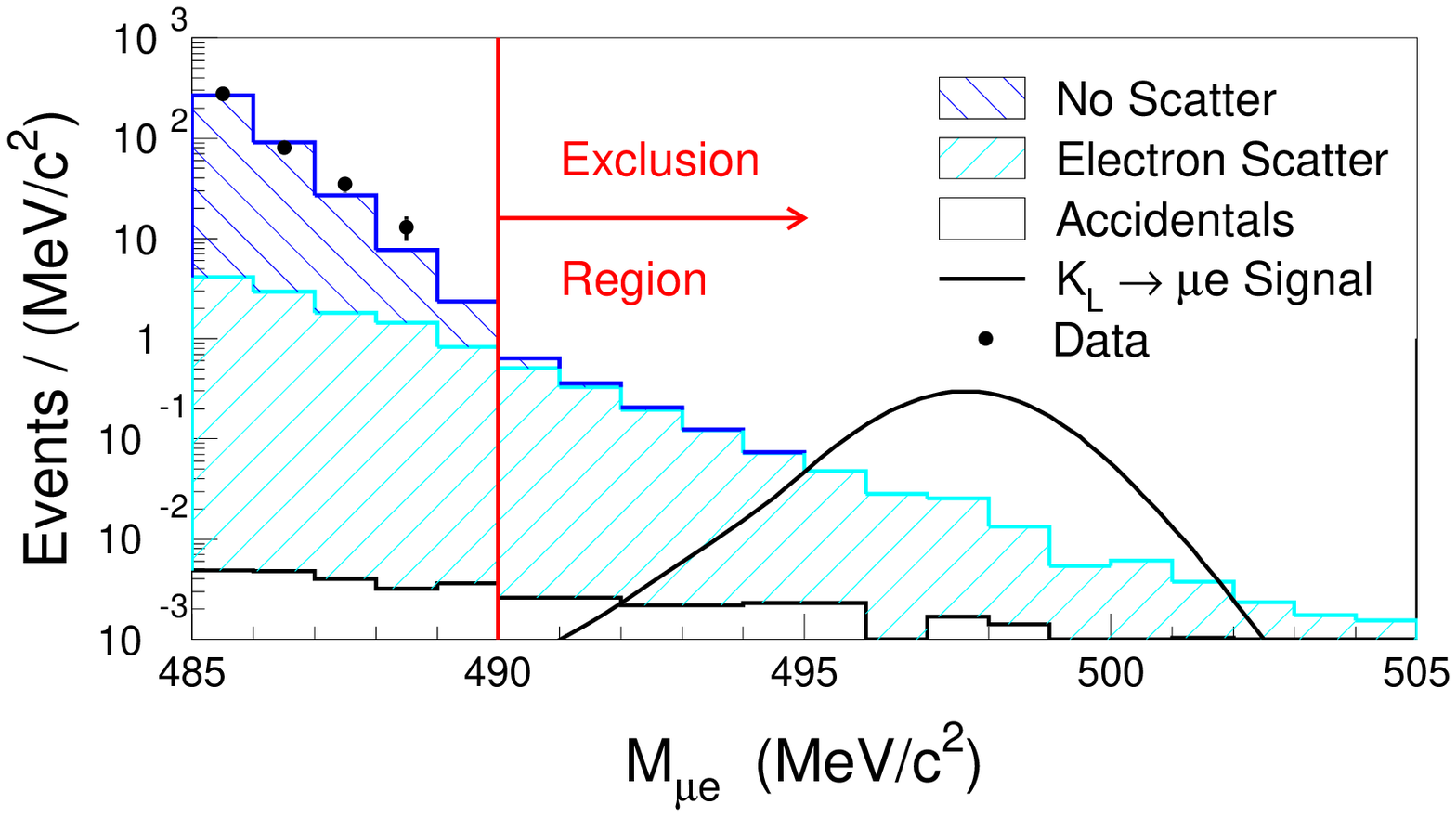,height=1.5in,clip=}}
\caption{On the left is a scatter plot of $\mathrm p_T^2$ versus $\mathrm M_{\mu e}$
from BNL E871.
The exclusion region for the blind analysis and the signal region are indicated by the
box and smaller contour, respectively. The plot on the right
shows the expected distributions for the signal and backgrounds, where the 
signal is shown for a branching fraction of $\mathrm 2 \times 10^{-12}$.
}
\label{kmueresult}
\end{center}
\end{figure}

To ensure that selection criteria were free of bias from knowledge
of potential signal events, they were chosen by studying
K$^0_L \rightarrow \mu^\pm e^\mp$ candidates 
outside an exclusion region larger than one that would  contain signal events 
(see Figure~\ref{kmueresult}).
Selection criteria included requirements on the quality of the tracks measured in the 
spectrometer and the existence of appropriate signals in 
the particle identification counters.
Large errors in M$_{\mu e}$ were shown by Monte Carlo simulation to come from 
$\mathrm K_L^0 \rightarrow \pi e \nu$ decays in which 
an electron scattered at large angles at or before the first tracking detector
and the pion decayed to a muon before the spectrometer. 
Accidental coincidences of two semi-leptonic decays were calculated to be a potential 
background; events with three or more fully reconstructed tracks in the spectrometer 
were rejected, reducing this background by an order of magnitude. 
The number and kinematic distributions of events with M$_{\mu e} >$ 493 MeV/c$^2$ 
were well reproduced by a Monte Carlo simulation of these processes; an example is 
shown in Figure~\ref{kmueresult}. 

Final selection criteria (including the choice of the
signal region) were optimized to reduce the expected background to 0.1 events and 
only then were events inside the exclusion region
analyzed. Figure~\ref{kmueresult} shows the final
distribution in $\mathrm p_T^2$ versus $\mathrm M_{\mu e}$. 
There are no events in the signal region and
the number of events in the exclusion region is consistent with the Monte Carlo 
prediction. Based on no observed events, a limit was set,
$\mathrm B(K^0_L \rightarrow \mu^\pm e^\mp) < 4.7 \times 10^{-12}$ [90\% CL]. 
This is the smallest limit ever set on a branching fraction for a hadron and 
results in a lower limit $\mathrm M_X > 190~ TeV/c^2$ assuming weak interaction
strength $\mathrm V \pm A$ coupling with maximal mixing.

\begin{figure}
\begin{center}
\hspace{.05in}
\hbox{\epsfig{figure=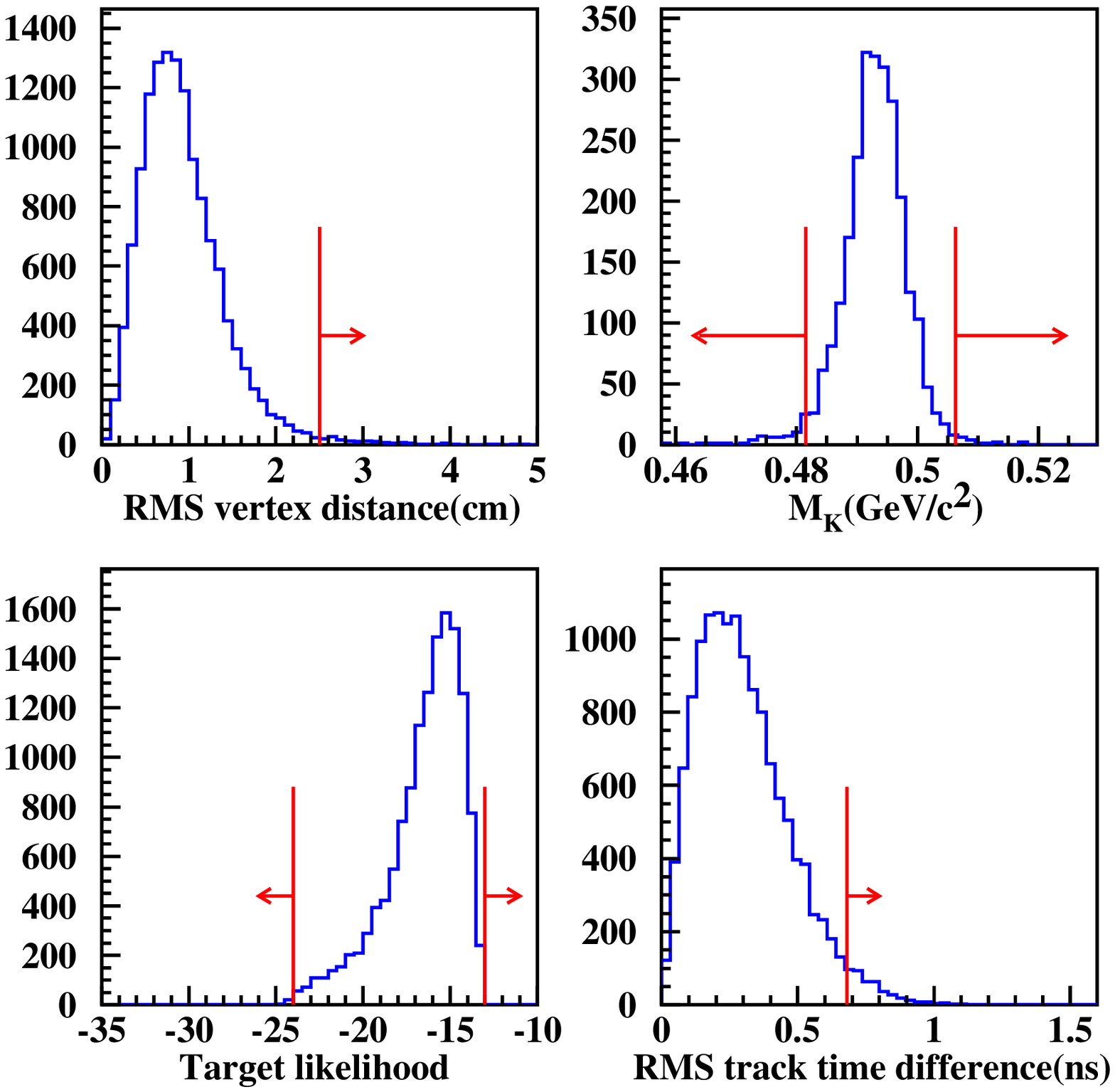,height=2.5in,clip=} 
\hspace{.5in} 
\epsfig{figure=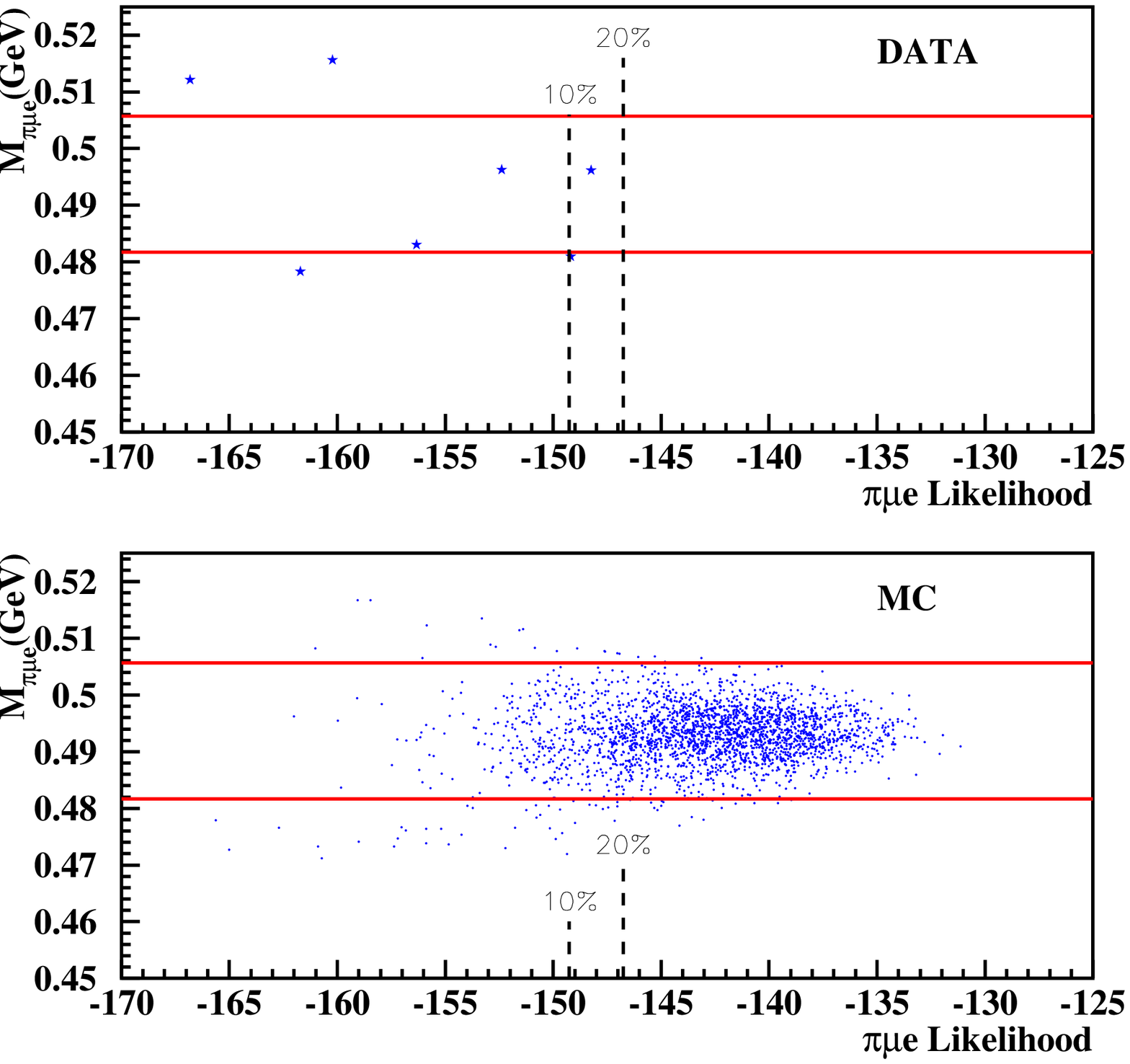,height=2.5in,clip=}}
\caption{On the left are distributions in four kinematic quantities used to select
$\mathrm K^+ \rightarrow \pi^+ \mu^+ e^-$ candidates, with the imposed cuts shown as
vertical lines. The scatter plot on the right is for $\mathrm M_{\pi^+ \mu^+ e^-}$ vs.
likelihood, for data and events simulated by Monte Carlo.}
\label{e865result}
\end{center}
\end{figure}

A search for the corresponding charged mode, $\mathrm K^+ \rightarrow \pi^+ \mu^+ e^-$,
was done by  BNL E865, which  used an intense, unseparated, 6 GeV/c negative beam.
In many respects the apparatus and detection techniques 
were similar to those of E871. K$^+$ mesons decayed in an evacuated
decay region. The decay products were first separated in a magnet, and then momentum 
analyzed using a proportional wire chamber spectrometer surrounding a second magnet. 
Cerenkov counters and a shashlik calorimeter identified e$^-$ on one side of the 
apparatus, and gas Cerenkov counters vetoed electrons on the other side, where 
$\mathrm \mu^+$ were also identified in a range stack. A series 
of selection criteria on kinematic quantities was imposed, as
shown in Figure~\ref{e865result}. The probability of a good
event having a particular value of the kinematic quantity was used to 
form a likelihood with which events were further selected. A scatter 
plot of the $\mathrm \pi^+ \mu^+ e^-$ mass vs. this likelihood is shown in 
Figure~\ref{e865result} for data and Monte Carlo generated 
$\mathrm K^+ \rightarrow \pi^+ \mu^+ e^-$ events.
Three events satisfied individual selection criteria but failed a 
likelihood requirement that was set (without prior knowledge of the distribution)
at a value that resulted in a 20\% loss of sensitivity.
Based on the absence of events satisfying all selection criteria, a limit
$\mathrm B(K^+ \rightarrow \pi^+ \mu^+ e^-) < 4.0 \times 10^{-11}$ was 
set~\cite{Zeller:1999}. 

These experiments may be the last to search for LFV using K decays. E871 has 
demonstrated the existence of a background at a sensitivity of order 
$\mathrm 10^{-13}$ resulting from $\mathrm K_L^0 \rightarrow
\pi e \nu$ decay with Mott scattering of the electron plus $\mathrm \pi$ decay 
conspiring to produce background indistinguishable from $\mathrm K_L^0 \rightarrow 
\mu^{\pm} e^{\mp}$ events; a 
dedicated experiment with a long data collection time would be required to reach 
that level. E865 is already at the point of being limited by background, and given
the inherently lower sensitivity of $\mathrm K^+ \rightarrow \pi^+ \mu^+ e^-$
in comparison with $\mathrm K_L^0 \rightarrow \mu^{\pm} e^{\mp}$, it is unlikely a
new effort will be undertaken. Much more sensitive probes of LFV are now being
proposed~\cite{Korenchenko:1998, Bachman:1997, RSVP:1999}  with muon processes, 
and this is likely where progress will be made.
\section{Quark Mixing Matrix Measurements}
The K system is one in which particularly incisive studies of the quark mixing (CKM) 
matrix can be made. Because FCNC decays are not allowed at tree level in the 
SM, many processes are dominated by one loop diagrams containing the top quark.
The best example of this is $\mathrm K_L^0 \rightarrow \pi^0 \nu \overline{\nu}$; 
it is pure CP violating (and predominantly {\it direct} CP violating), and the 
rate can be unambiguously related to the parameter $\mathrm \eta$ in the Wolfenstein 
parameterization of the CKM matrix. 
There are also experimental advantages to doing experiments with kaons. Because
of  their long lifetime, clean, intense beams can be produced, often with little 
contamination from other particles. Experiments can 
be done far from the environment in which the kaons are produced, allowing for 
experiments with a branching fraction sensitivity that cannot be contemplated in 
a collider environment. 

The experiments to study the CKM matrix focus on decays of charged or neutral kaons
to pairs of leptons, with or without an additional pion. The most important are
$\mathrm K_L^0 \rightarrow \pi^0 \nu \overline{\nu}$,  
$\mathrm K^+ \rightarrow \pi^+ \nu \overline{\nu}$, 
$\mathrm K_L^0 \rightarrow \mu^+ \mu^-$, 
$\mathrm K_L^0 \rightarrow \pi^0 e^+ e^-$ and  
$\mathrm K_L^0 \rightarrow \pi^0 \mu^+ \mu^-$.
Other modes (for example $\mathrm K_L^0 \rightarrow e^+ e^-$ and 
$\mathrm K^+ \rightarrow \pi^+ {\mathit l^+ l^-}$) are dominated by radiative decay processes and 
will not have much input on CKM measurements. 
In the remainder of this section, I will briefly review the phenomenology of
K decays and the CKM matrix, and then discuss recent results. 
\subsection{Phenomenology of CKM Matrix and 
$\mathbf K \rightarrow (\pi) {\mathit l \overline{l}}$} 
\begin{figure}[b]
\begin{center}
\centerline{\hbox{\hspace{-.05in}\psfig{figure=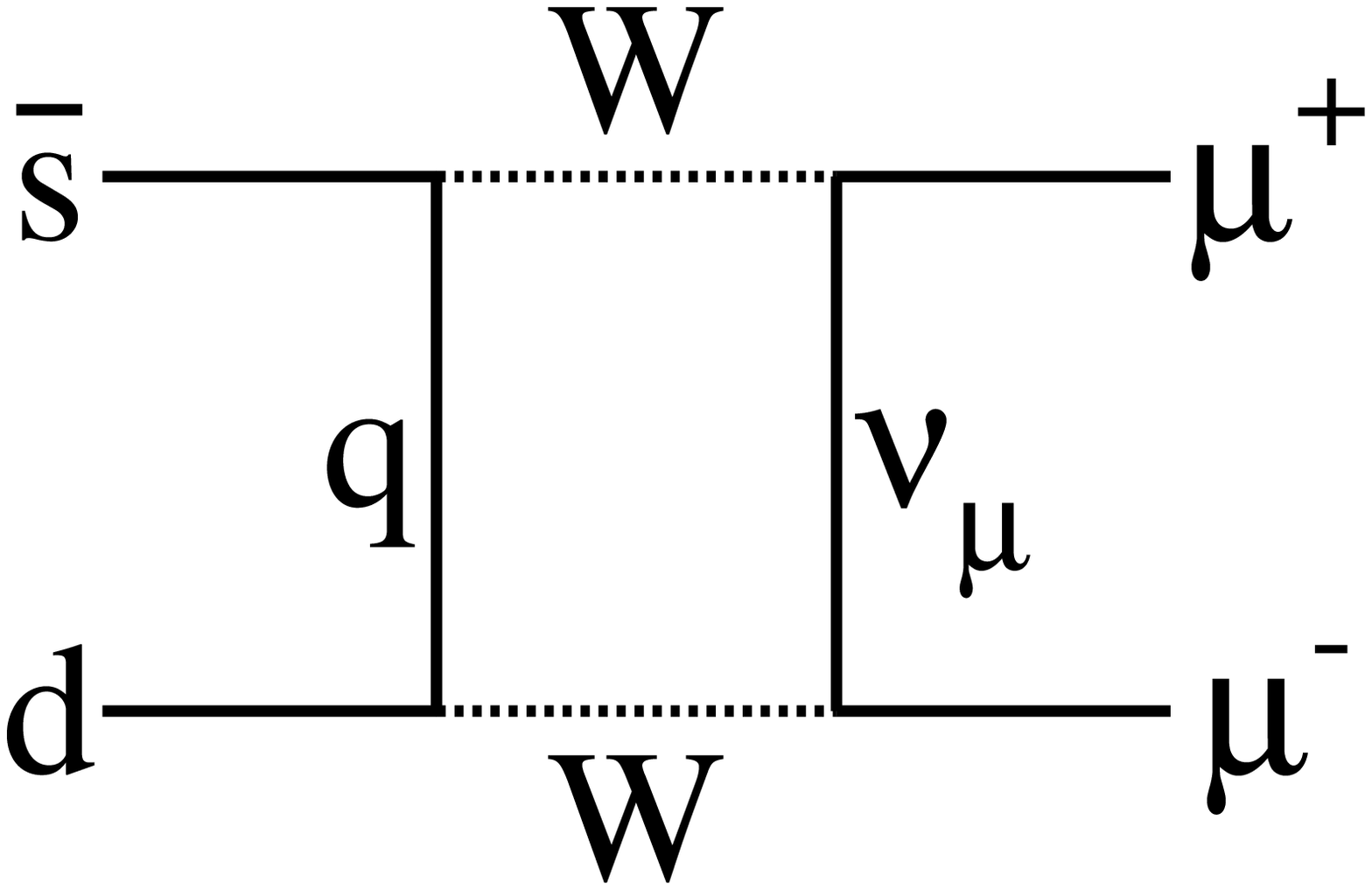,width=.9in} 
\hspace{.3in} \psfig{figure=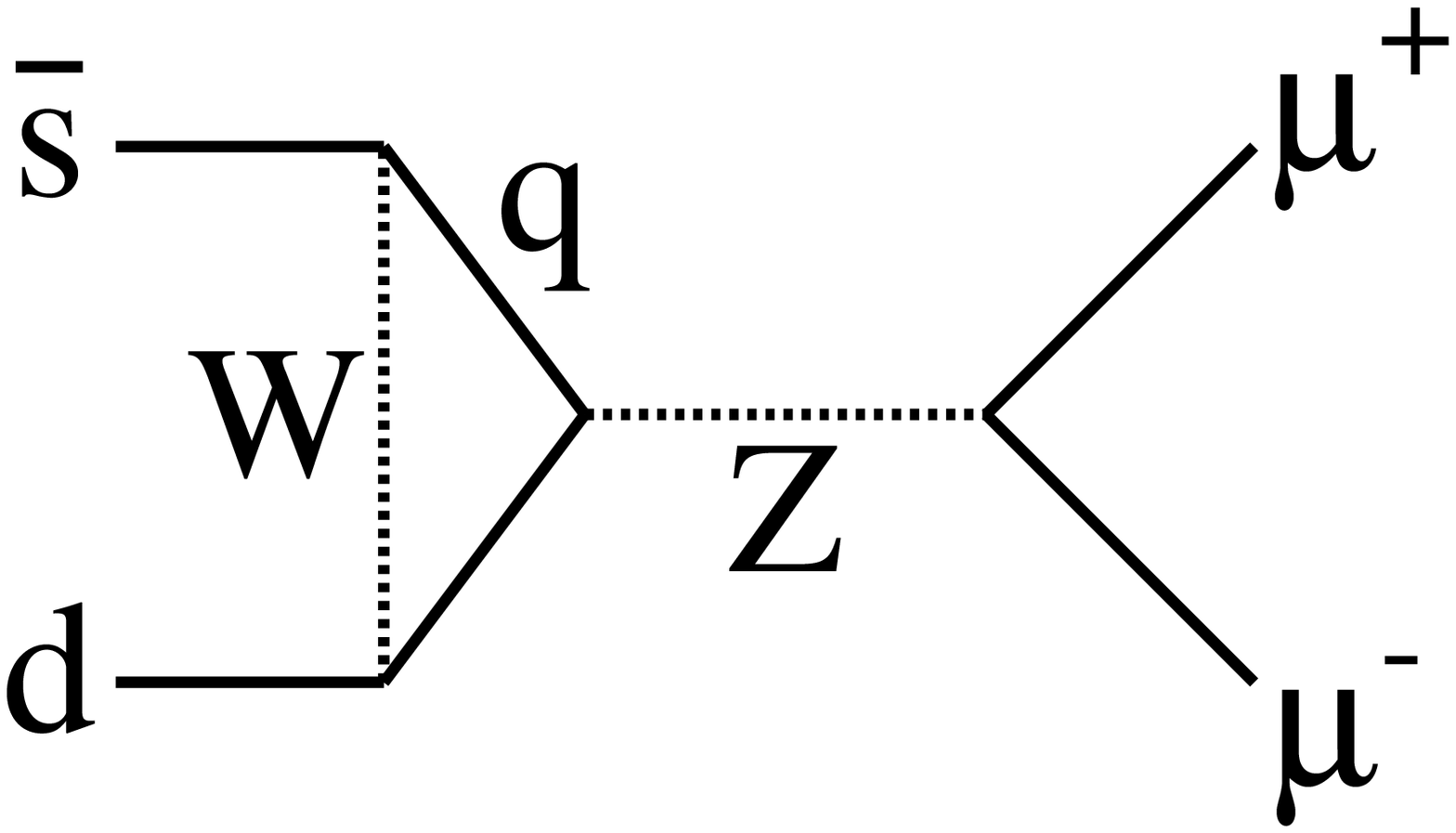,width=.9in} 
\hspace{.3in} \psfig{figure=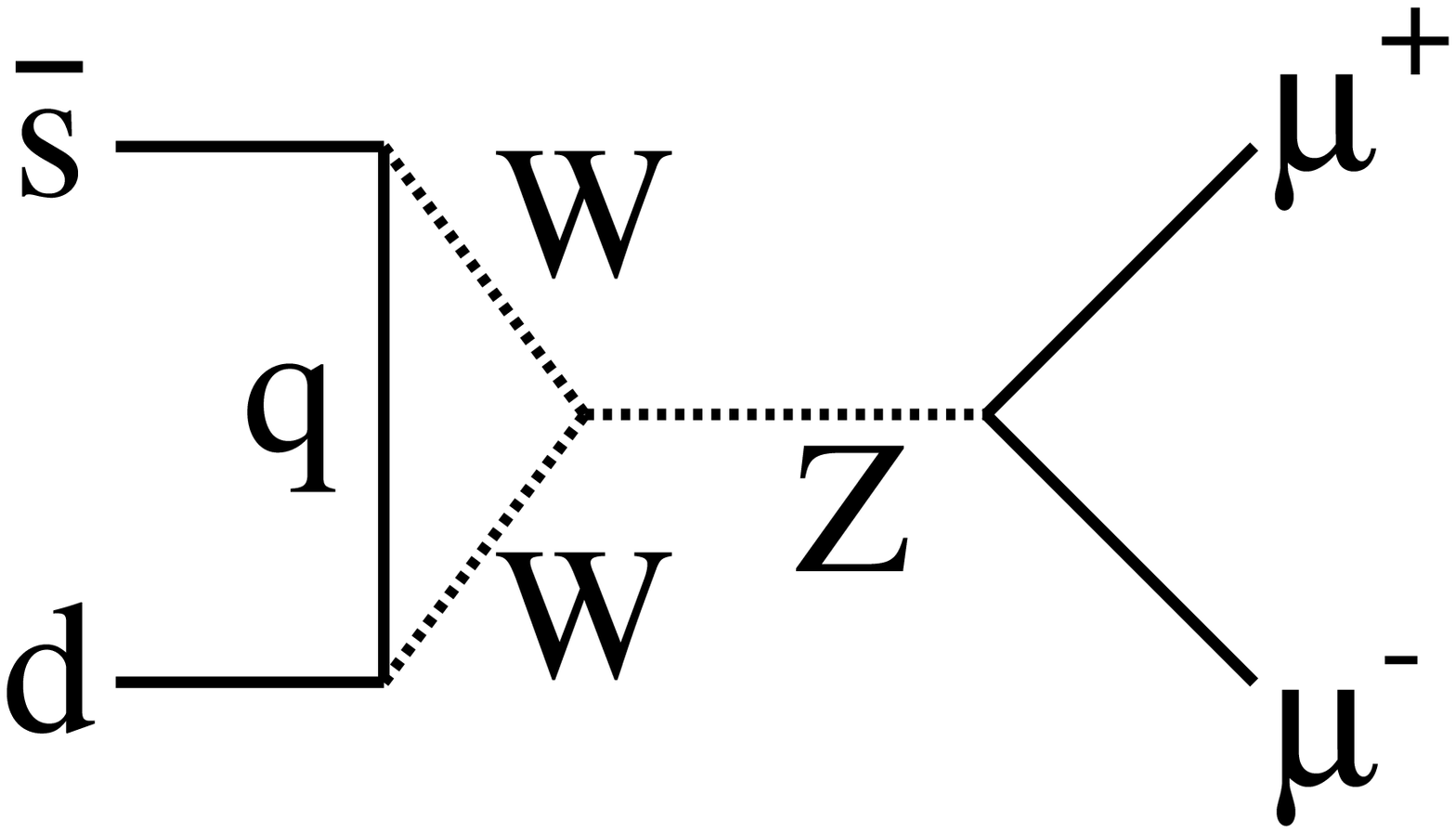,width=.9in}
\hspace{.4in}\psfig{figure=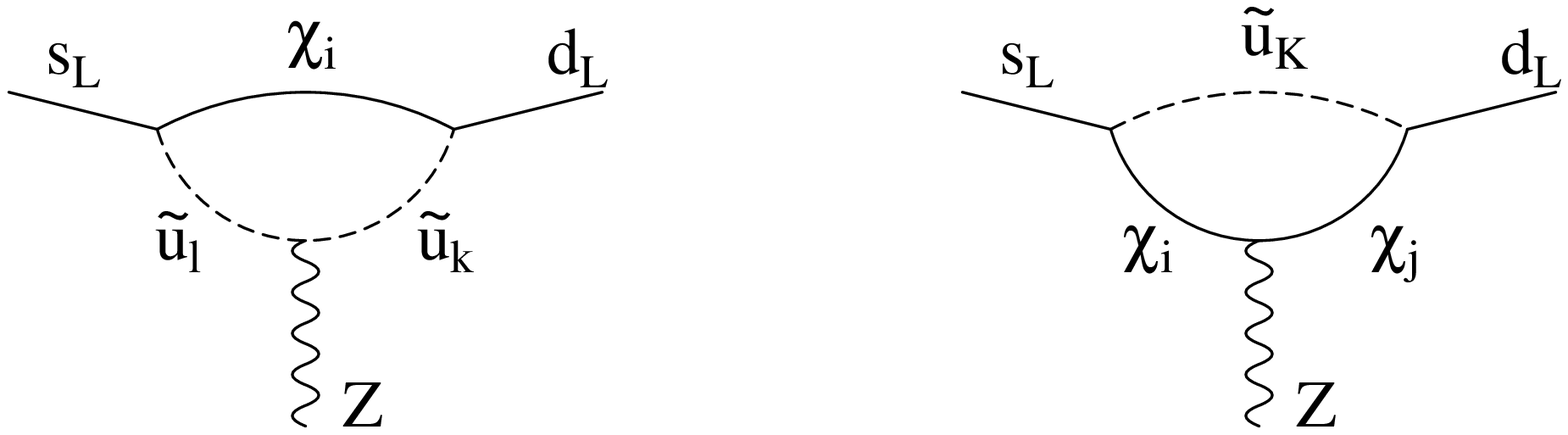,height=.65in,clip=}}}
\caption{The three diagrams  on the left are SM box and penguin diagrams 
contributing to  $\mathrm K \rightarrow (\pi) l \overline{l}$  decays, for example
$\mathrm K^0 \rightarrow \mu^+ \mu^-$. The two diagrams on the right are  
typical of those that arises in supersymmetry models and that may
also contribute to these decays and to indirect CP violation.}
\label{kpilldiagrams}
\end{center}
\end{figure}

Within the context of the SM, the rates for these rare decays can be calculated
with small theoretical uncertainty. They proceed through box and penguin
diagrams of the type shown in Figure~\ref{kpilldiagrams}. They are dominated
by the top quark contribution and depend on the CKM matrix elements 
$\mathrm V_{ts}$ and  $\mathrm V_{td}$. 
The short distance contributions to the decay amplitudes are proportional 
to $\mathrm Re(V_{td}V_{ts}^*) = A^2 \lambda^5(1-\rho)$ for \KLMUMU, to 
$\mathrm Im(V_{td}V_{ts}^*) = A^2 \lambda^5 \eta$ for \KLPINUNU, and to 
$\mathrm V_{td}V_{ts}^* = A^2 \lambda^5 (1-\rho-i\eta)$ for \KPPINUNU. 
For the latter two, any additional complications in the calculation of the 
decay rate are negligible; hadronic effects
are incorporated by comparing to corresponding semi-leptonic decays, and contributions
from charm quark diagrams are small (and only relevant for 
$\mathrm K^+ \rightarrow \pi^+ \nu \overline{\nu}$, since 
$\mathrm K_L^0 \rightarrow \pi^0 \nu \overline{\nu}$ is purely CP violating). 
In the case of $\mathrm K_L^0 \rightarrow \mu^+ \mu^-$ there are significant 
long distance effects (primarily from $\mathrm 2 \gamma$ intermediate states) 
that very much complicate the extraction of short distance physics.

As was originally pointed out by Jarlskog~\cite{Jarlskog:1985} 
in the context of the CKM matrix, all triangles in the complex plane formed by the 
inner product of pairs of rows or columns of a unitary matrix have the 
same area and this area is proportional to $\mathrm \eta$ and hence to the 
degree of CP violation in the SM. To order $\mathrm \lambda^6$, the area is given by 
$\mathrm A = J/2 = \frac{1}{2} \times \lambda(1-\lambda^2/2) \times A^2 \lambda^5 \eta$;
J is referred to as the Jarlskog invariant. 
The triangle formed by the inner product of the first and third columns is often 
referred to as the {\it unitarity triangle}. As has been stressed recently by a number of
authors, any triangle is equally good (from a theoretical viewpoint) in quantifying
CP violation in the CKM matrix.
K decays are particularly powerful in measuring J using the first and second 
columns to form a triangle. The base then depends only on the Cabibo angle, and the 
height is proportional to $\mathrm B(K_L^0 \rightarrow \pi^0 \nu \overline{\nu})$.
Measurements of $\mathrm B(K^+ \rightarrow \pi^+ \nu \overline{\nu})$  and  
$\mathrm B(K_L^0 \rightarrow \mu^+ \mu^-)$ provide constraints 
on the closure of the triangle and might  indicate deviations from the SM prediction. 
It is more difficult to determine J from B decays, but eventually the value determined
in that system could be compared with that derived from K decays to again test the SM.

There has recently been renewed interest in K decay measurements from two sources.
First, E787 at BNL reported~\cite{Adler:1997} the first observation of the 
decay $\mathrm K^+ \rightarrow \pi^+ \nu \overline{\nu}$, with one event corresponding 
to a branching fraction of $\mathrm 4.2^{+9.7}_{-3.5} \times 10^{-10}$. 
This is about a factor of five
above the SM prediction. Second, the KTeV~\cite{Alavi-Harati:1999} and 
NA48~\cite{Fanti:1999} collaborations have presented first results
of their new measurements of $\mathrm \epsilon^{\prime}/\epsilon$. They are 
in good agreement
with an earlier result~\cite{Barr:1993} from NA31 at CERN and somewhat higher 
than the result~\cite{Gibbons:1993} from E731 at Fermilab. The value is significantly
different from zero and also significantly higher than most expectations. Although the
recent results on  $\mathrm K^+ \rightarrow \pi^+ \nu \overline{\nu}$ and 
$\mathrm \epsilon^{\prime}/\epsilon$ hint at the possibility of new physics 
contributions, neither requires it.
The SM prediction  of the $\mathrm K^+ \rightarrow \pi^+ \nu \overline{\nu}$ rate
is theoretically unambiguous, but the measurement is not statistically inconsistent 
with it. On the other hand,
the average of the $\mathrm \epsilon^{\prime}/\epsilon$ measurements is significantly 
different from most predictions, but the model calculations are theoretically 
ambiguous. 

Given these hints, three questions arise 
in understanding the extent to which experiments can make a definitive statement 
about the need for non-SM physics.
First, is the SM predictions for $\mathrm \epsilon^{\prime}/\epsilon$ currently 
inconsistent with experiment and can it be made reliable?
Second, what new physics could contribute to 
$\mathrm \epsilon^{\prime}/\epsilon$ and the decay 
$\mathrm K^+ \rightarrow \pi^+ \nu \overline{\nu}$? Third, do measurements of rare 
decays of kaons limit possible contributions from non-SM sources and can they provide 
unambiguous evidence for deviations from the SM?

The first question remains controversial. Bosch et al.~\cite{Bosch:1999}
have varied all parameters entering into the SM prediction for 
$\mathrm \epsilon^{\prime}/\epsilon$ and contend that the experimental result can be 
accommodated within the SM only by simultaneously stretching many of the values entering
into the prediction, including the values of matrix elements derived from either
lattice calculations or phenomenological models. The second question has been 
studied in the context of effective Zds couplings of the type shown in 
Figure~\ref{kpilldiagrams}. These arise naturally in 
supersymmetric theories and could easily give large contributions to both 
$\mathrm \epsilon^{\prime}/\epsilon$  and rates for rare K decays. 
For example, Bosch et al.~\cite{Bosch:1999} infer a SM prediction of  
$\mathrm 1.6 \times 10^{-11} < B(K_L^0 \rightarrow \pi^0 \nu \overline{\nu}) 
< 3.9 \times 10^{-11}$ and show that 
$\mathrm \epsilon^{\prime}/\epsilon < 28 \times 10^{-4}$ implies 
$\mathrm B(K_L^0 \rightarrow \pi^0 \nu \overline{\nu}) < 48 \times 10^{-11}$ 
if the enhanced value of $\mathrm \epsilon^{\prime}/\epsilon$ is due to a new 
effective Zds coupling. Within this context, effective Zds couplings with large 
contributions to $\mathrm B(K_L^0 \rightarrow \pi^0 \nu \overline{\nu})$
 are not excluded. The advantage of a measurement of this decay rate 
is that the SM prediction is unambiguous within a rather narrow window and 
deviations would provide definitive evidence for non-SM physics that may not
obtain from even a very precise measurement of $\mathrm \epsilon^{\prime}/\epsilon$. 
Similarly, SM physics and current measurements imply~\cite{Buras:1999}
$\mathrm B(K^+ \rightarrow \pi^+ \nu \overline{\nu} = 8.2 \times 10^{-11}$, 
whereas the above limit on $\mathrm \epsilon^{\prime}/\epsilon$
plus the measured branching fraction for $\mathrm K_L^0 \rightarrow \mu^+ \mu^-$ implies 
$\mathrm B(K^+ \rightarrow \pi^+ \nu \overline{\nu}) < 29 \times 10^{-11}$ 
even with arbitrary effective Zds couplings. 
Again, current measurements allow significant contributions to the  
$\mathrm K^+ \rightarrow \pi^+ \nu \overline{\nu}$ decay rate from non-SM physics, 
and a precise measurement of this rate will restrict this possibility. I next discuss 
recent experimental results that bear on these questions.
\subsection{Measurement of $\mathbf B(K_L^0 \rightarrow {\mathit l^+ l^-})$}
The decay \KLMUMU is by far the best measured of the modes discussed. 
Recently, E871 at BNL has reported final results~\cite{Ambrose1:1998, Ambrose:1999}
of measurements of B($\mathrm K_L^0 \rightarrow e^+ e^-)$  
and B($\mathrm K_L^0 \rightarrow \mu^+ \mu^-)$.
The experiment used the same apparatus as that used for the 
$\mathrm K_L^0 \rightarrow \mu^{\pm} e^{\mp}$ search discussed earlier 
and data were collected during the same period. In the analysis, 
the emphasis was not on eliminating background but rather on understanding the 
relative acceptance for $\mathrm K_L^0 \rightarrow \mu^+ \mu^-$ decays and the  
$\mathrm K_L^0 \rightarrow \pi^+ \pi^-$ decays
to which the experiment was normalized. To avoid biases in the analysis, an 
unknown prescale factor was applied to the normalization sample and only revealed 
and corrected for after the analysis was completed. 

\begin{figure}[htb]
\begin{center}
\hbox{\hspace{.3in}
\psfig{figure=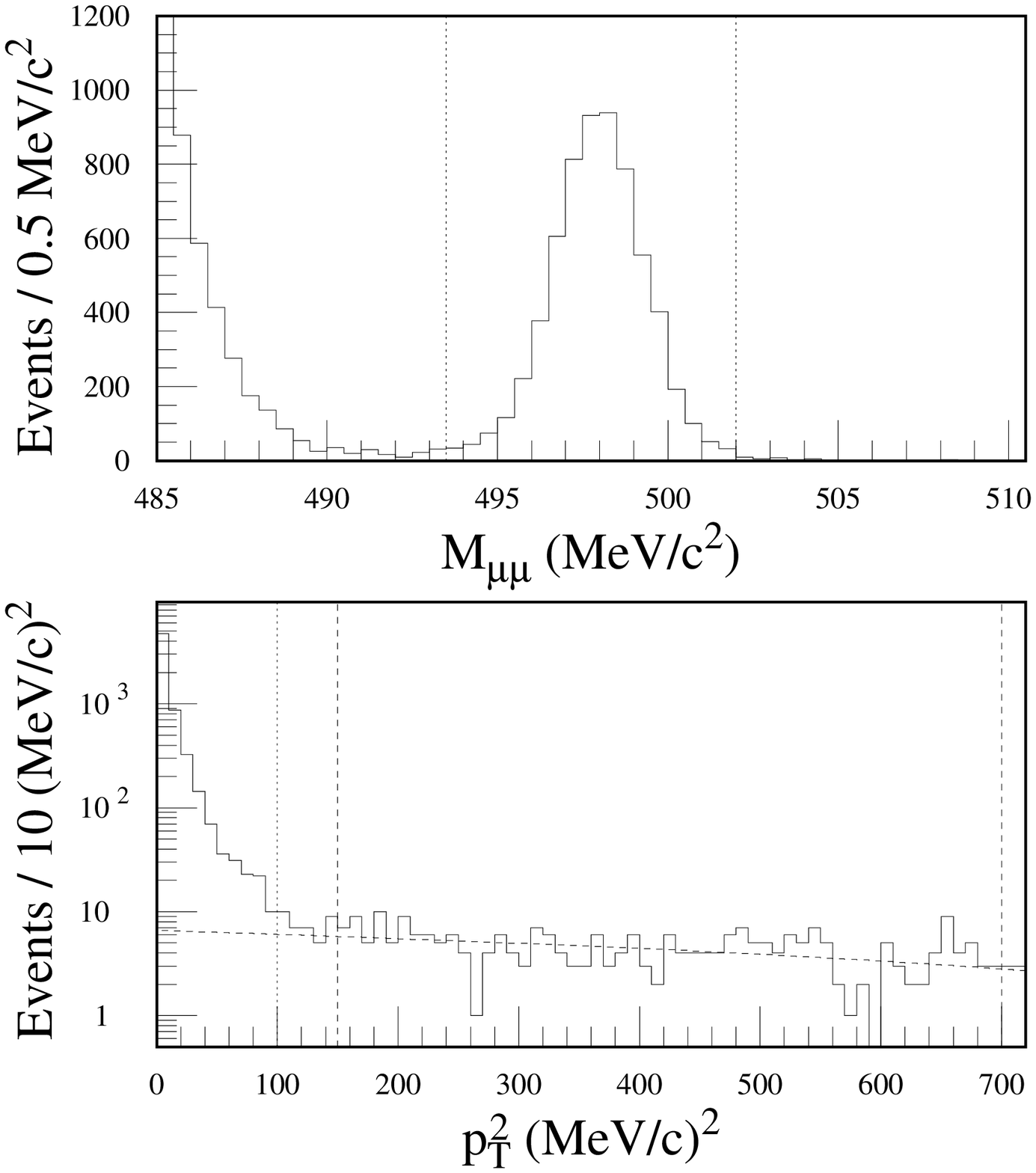,height=2.in,width=2.5in,clip=}
\hspace{.2in}
\epsfig{figure=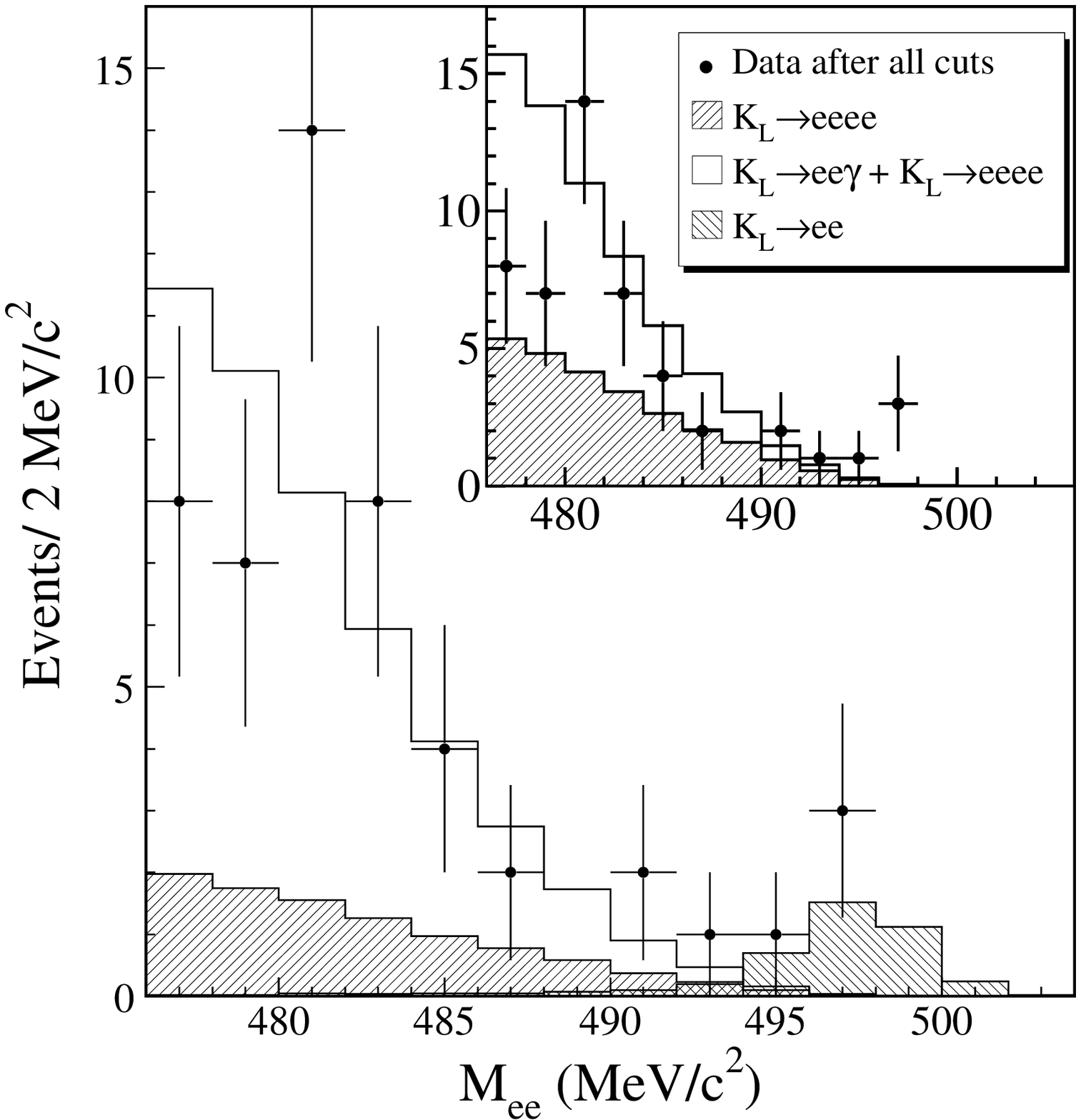,height=2.in,width=2.5in,clip=}  }
\caption{The plots show the $\mu \mu$ mass plot and the distribution in 
$\mathrm p_T^2$, the latter with a linear fit to the background superimposed.}
\label{e871klmumu}
\end{center}
\end{figure}

Figure~\ref{e871klmumu} shows distributions in the $\mu^+\mu^-$ 
invariant mass and square of the transverse momentum for events satisfying 
appropriate selection criteria.
A small ($\sim$1.2\%) background in the $\mathrm K_L^0 \rightarrow \mu^+ \mu^-$ 
candidate event sample (primarily from $\mathrm K_L^0 \rightarrow \pi e \nu$
events)
was subtracted by linearly extrapolating the level of events at large $\mu^+\mu^-$ 
transverse momentum to under the $\mathrm K_L^0 \rightarrow \mu^+ \mu^-$ peak. 
The $\mathrm K_L^0 \rightarrow \mu^+ \mu^-$ sample contains 
$\sim$6200 events. Appropriate correction factors
(for example for trigger and particle identification efficiencies)
were applied to the relative $\mathrm \mu^+\mu^-$ and 
$\mathrm \pi^+\pi^-$ acceptances; these were
typically derived from ancillary measurements and were typically a few percent.
The largest of these were for pion absorption in the spectrometer (5.2\%), the online 
$\mu^+\mu^-$ trigger efficiency ($\sim$3\%), and the off-line $\mu^+\mu^-$ particle
identification efficiency ($\sim$6\%). The resulting branching ratio is 
$\mathrm 
\frac{\Gamma(K_L^0 \rightarrow \mu^+ \mu^-)}{\Gamma(K_L^0 \rightarrow \pi^+ \pi^-)}
= (3.474 \pm 0.057) \times 10^{-6}$; multiplying by the measured value of 
$\mathrm B(K_L^0 \rightarrow \pi^+ \pi^-)$ results in 
$\mathrm B(K_L^0 \rightarrow \mu^+ \mu^-) = (7.18 \pm 0.17) \times 10^{-9}$. 
Statistical and systematic errors are added in quadrature; 
the largest of these are the statistical uncertainty on the number of 
$\mathrm K_L^0 \rightarrow \mu^+ \mu^-$ events (1.32\%) and 
systematic uncertainties on pion absorption (0.50\%) and the  relative geometrical and 
reconstruction efficiencies (0.57\%).

An estimate of the short distance contribution to this decay is derived by 
first subtracting the ``unitarity bound'' due to the absorptive 2$\gamma$ contribution 
to the decay rate, derived from the measured $K_L^0 \rightarrow \gamma \gamma$
decay rate and a QED calculation. This results in a measured dispersive part of 
the branching fraction of $\mathrm (0.11 \pm 0.18) \times 10^{-9}$ or 
$\mathrm |ReA_{exp}|^2 < 0.37 \times 10^{-9}$ [90\% CL]. Proceeding beyond this requires 
a model for calculating the long distance dispersive amplitude, a controversial 
procedure~\cite{Valencia:1998}. Using one recent estimate~\cite{D'Ambrosio:1998}, the 
Wolfenstein parameter $\rho$ is found to be bounded from below at $-0.33$, 
consistent with existing constraints.  

E871 has also measured~\cite{Ambrose1:1998} the branching fraction for the decay 
\KLEE for the first time. This decay rate is suppressed with respect to that of 
$\mathrm K_L^0 \rightarrow \mu^+ \mu^-$ due to helicity 
considerations and is expected to be completely dominated by long distance effects. 
Unlike the case of $\mathrm K_L^0 \rightarrow \mu^+ \mu^-$, these can be reliably 
calculated~\cite{Valencia:1998,Dumm:1998} and the 
branching fraction is expected to be $\sim 10^{-11}$. The analysis proceeded in much 
the same way as for $\mathrm K_L^0 \rightarrow \mu^+ \mu^-$. Backgrounds resulted primarily from physics sources:
$\mathrm K_L^0 \rightarrow e^+ e^- \gamma$ and 
$\mathrm K_L^0 \rightarrow e^+ e^- e^+e^-$ decays. 
Contributions from the latter were minimized by removing events in which extra 
tracks pointing to the decay vertex were detected before the first magnet. 
Figure~\ref{e871klmumu} shows the $\mathrm e^+ e^-$ mass distribution and a fit
to \KLEE signal plus background from $\mathrm K_L^0 \rightarrow e^+ e^- \gamma$ and 
$\mathrm K_L^0 \rightarrow e^+ e^- e^+e^-$. The fit yields 4 $\mathrm K_L^0 
\rightarrow e^+ e^-$ events and a corresponding value of 
$\mathrm B(K_L^0 \rightarrow e^+e^-) = (8.9^{+4.5}_{-2.8}) \times 10^{-12}$. 
This is in good agreement with predictions of chiral perturbation theory calculations
and is the smallest branching fraction ever measured. 

These measurements of $\mathrm B(K_L^0 \rightarrow e^+e^-)$ and  
$\mathrm B(K_L^0 \rightarrow \mu^+\mu^-)$ are unlikely to be improved upon
in the foreseeable future. For the former, little can be learned about short distance 
physics from improved measurements. For the latter, extracting a value for $\mathrm \rho$
is now limited by model assumptions in the calculation of the long distance dispersive
contributions, and until they can be improved, better precision on this branching 
fraction measurement is not useful. 

\subsection{Measurement of $\mathbf B(K^+ \rightarrow \pi^+ \nu \overline{\nu})$}
The measurement of B($\mathrm K^+ \rightarrow \pi^+ \nu \overline{\nu}$) allows a 
direct determination of the magnitude of $\mathrm V_{td}V_{ts}^*$ (after a small 
correction for c quark contribution); hence it measures directly the magnitude 
of one side of the triangle formed by $\mathrm V_{id}V_{is}^*$. 
Experimental difficulties arise
from the facts that the 2 neutrinos are unobservable and that many $\pi^+$ are also 
produced from the copious decay mode $\mathrm K^+ \rightarrow \pi^+ \pi^0$. 
E787 at BNL detected this decay for the first time~\cite{Adler:1997}, 
observing one event with a branching fraction consistent with the SM prediction. 
The event was found first in a data sample corresponding to a sensitivity of 
about 0.2 events at the SM level, raising the hope that there might be a non-SM
contribution, despite the high probability of finding an event at that sensitivity.
Nonetheless, this hope motivated some useful theoretical discussion of what might 
be responsible for a rate larger than that predicted by the SM. 

\begin{figure}[htb]
\begin{center}
\hbox{\hspace{.5in}
\epsfig{figure=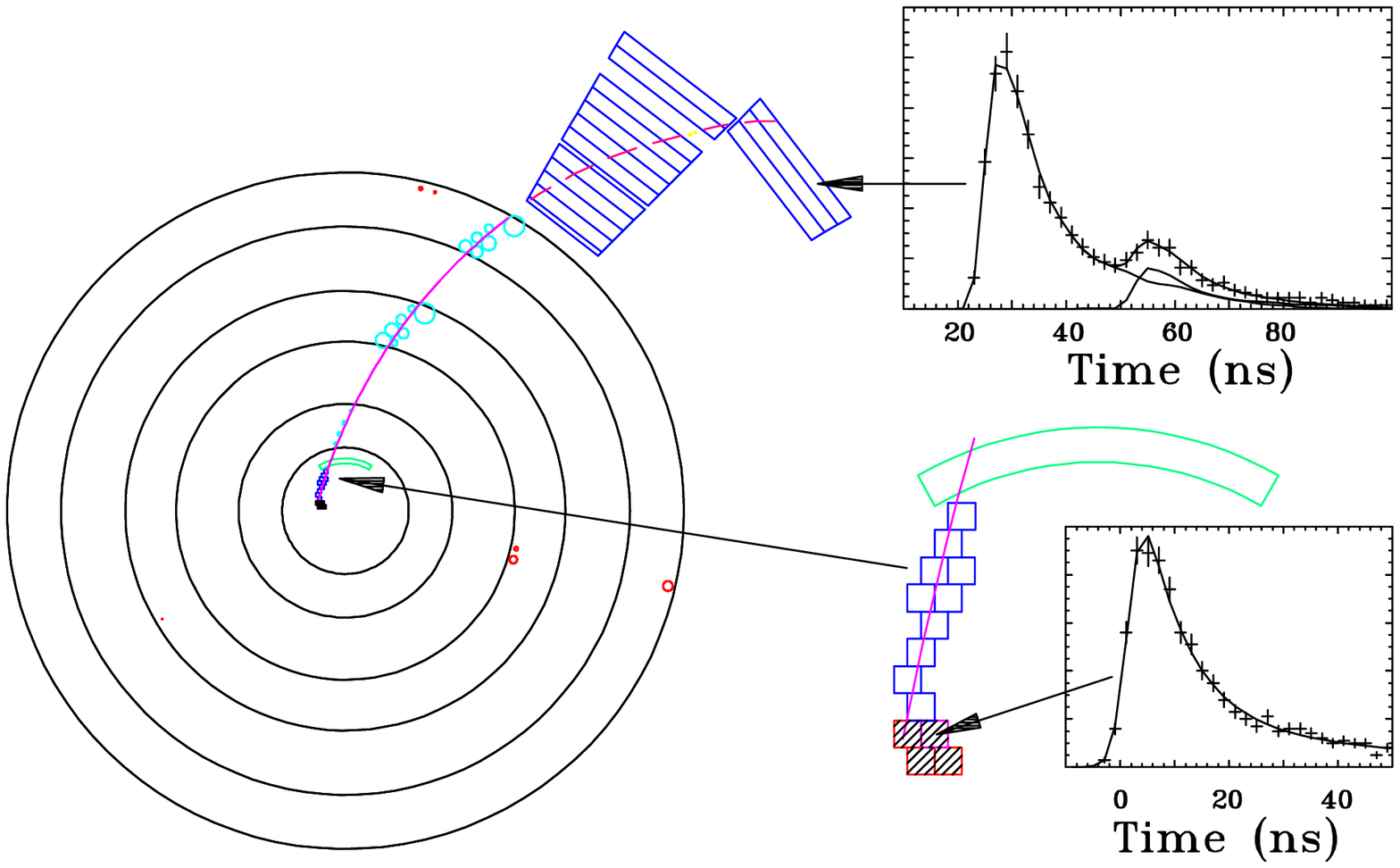,height=1.5in,clip=} \hspace{.5in}

  \epsfig{figure=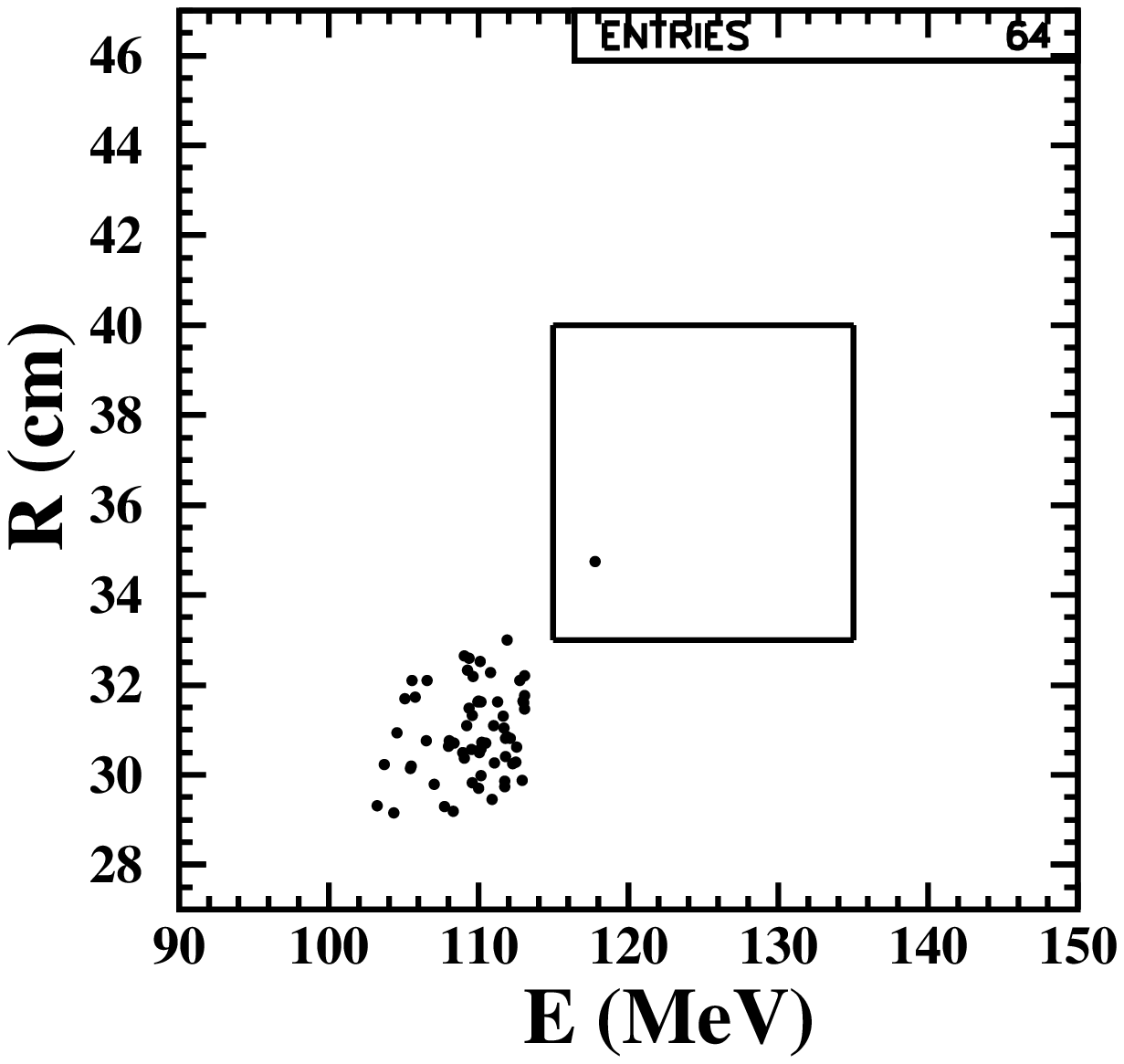,height=1.5in,clip=}  }
\caption{The figure shows a scatter plot of range vs. energy for candidate 
$\mathrm K^+ \rightarrow \pi^+ \nu \overline{\nu}$ events following event selection on the left and an event display of the 
one signal event on the right.}
\label{e787}
\end{center}
\end{figure}
E787 stops $\mathrm K^+$ mesons in an active target and then 
measures all conceivable final state properties: the $\pi^+$ kinetic
energy (determined from the energy deposited in the target and in a plastic 
scintillator range stack), the momentum (measured in a magnetic spectrometer), 
the range (determined from the penetration distance in the range stack), and 
the decay chain $\mathrm \pi \rightarrow \mu \rightarrow e$. An event display 
of the one candidate is shown in Figure~\ref{e787}.  
The figure also shows a scatter plot of  the range vs. energy for events 
satisfying a set of selection criteria on decay chain, momentum, energy in photon
veto counters, and tracking quality. About 70\% of 
$\mathrm K^+ \rightarrow \pi^+ \nu \overline{\nu}$ events satisfying these criteria
would fall in the rectangular box. Background from 
$\mathrm K^+ \rightarrow \mu^+ \nu_{\mu}$ would be at larger range and energy. 

E787 recently presented preliminary 
results~\cite{Kettell:1999} from a larger data set, 
with no additional events seen. Based on the total sensitivity currently reported, 
the branching fraction is given by   
$\mathrm B( K^+ \rightarrow \pi^+ \nu \overline{\nu})  = (1.5^{+3.5}_{-1.3}) 
\times 10^{-10}$ [90\% CL]. 

Improvements in the precision of this measurement will come in stages. Data currently exists
to improve the sensitivity by about a factor of two, with analysis expected to be 
completed in about 1 year. A modestly upgraded experiment, E949~\cite{E949:1998}, 
was recently approved  at BNL, and the U.S. DOE has agreed to 
operate the accelerator for this experiment for about 5000 additional hours of running. 
It is expected to get 7-14 events at the SM level. 
Further improvement in sensitivity may come from the CKM experiment~\cite{CKM:1998} 
proposed at Fermilab to detect about 100 events at the SM branching fraction. This 
experiment uses a technique completely different from that of E949, using 
$\mathrm K^+$ decay in flight from a separated 12 GeV beam.

\subsection{Search for $\mathbf K_L^0 \rightarrow \pi^0 \nu \overline{\nu}$}
Despite the importance of a measurement of the branching fraction for the 
decay $\mathrm K_L^0 \rightarrow \pi^0 \nu \overline{\nu}$, progress to date has 
come only as a by-product of other experiments, primarily from the KTeV 
collaboration. Recently, this group has published results based on two techniques.
They are distinguished by the detection of $\pi^0$ decays to either 
$\mathrm \gamma \gamma$ or $\mathrm \gamma e^+ e^-$.
The experimental difficulties in the experiment result from the lack of knowledge of
the initial state (since the beam is neither mono-energetic nor very small) and
the paucity of information about the final state. For the 2$\gamma$ decay mode, only 
the energy and position of the two photons is measured, and the $\mathrm K_L^0$ 
decay point is inferred by constraining the $\mathrm \gamma \gamma$ 
invariant mass 
to the $\mathrm \pi^0$ mass. For Dalitz decays, the $\mathrm K_L^0$ decay 
point is determined by the vertex position of the $\mathrm e^+e^-$ pair. In both 
cases, copious backgrounds exist from $\mathrm K_L^0 \rightarrow \pi^0 \pi^0$ and
$\mathrm K_L^0 \rightarrow \pi^0 \pi^0 \pi^0$ decays and from hyperon decays.

\begin{figure}[htb]
\epsfig{figure=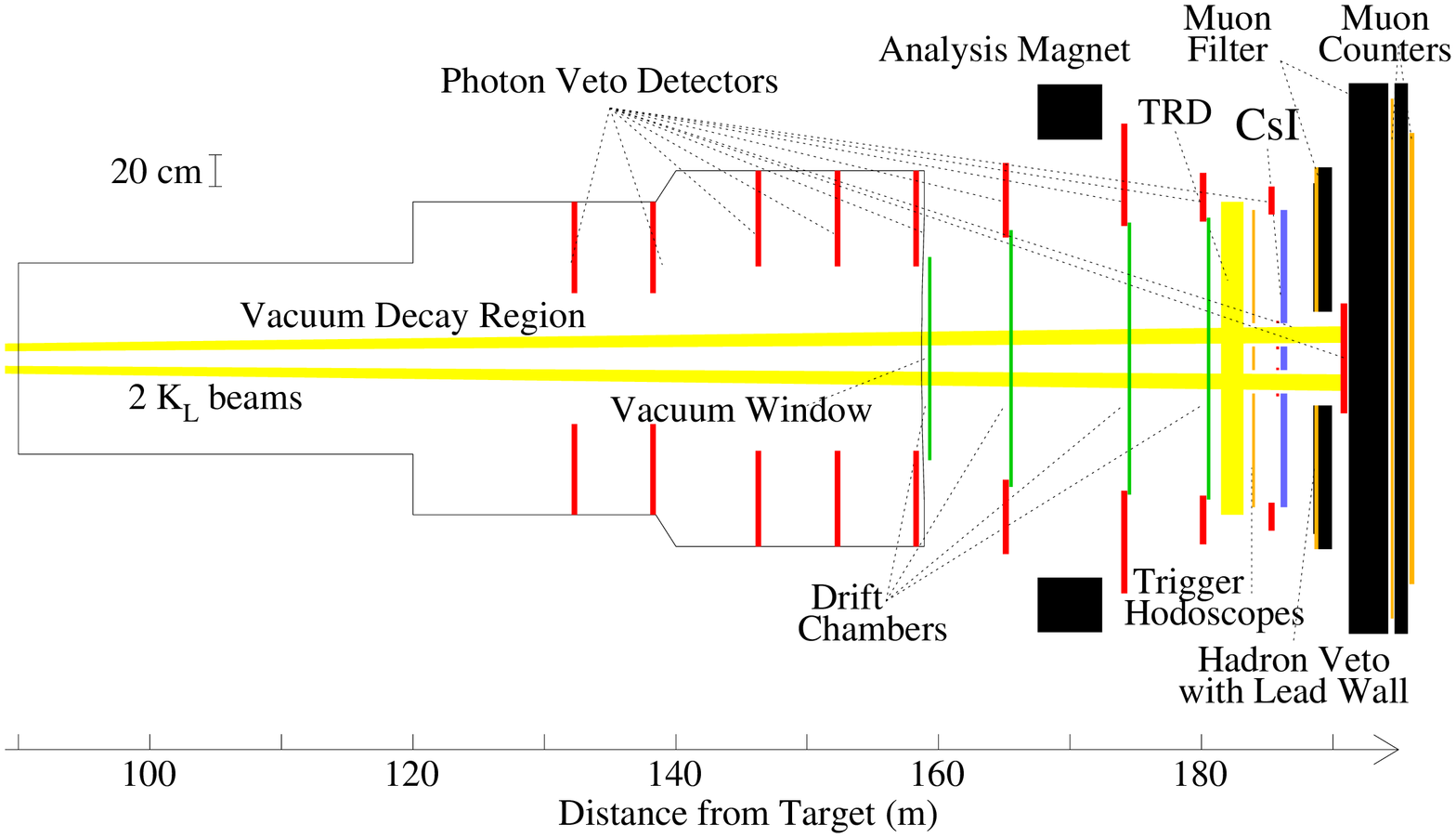,height=3.5in,angle=-90,clip=}

\vspace{-1.9in}
\hbox{\hspace{4.1in} 
\epsfig{figure=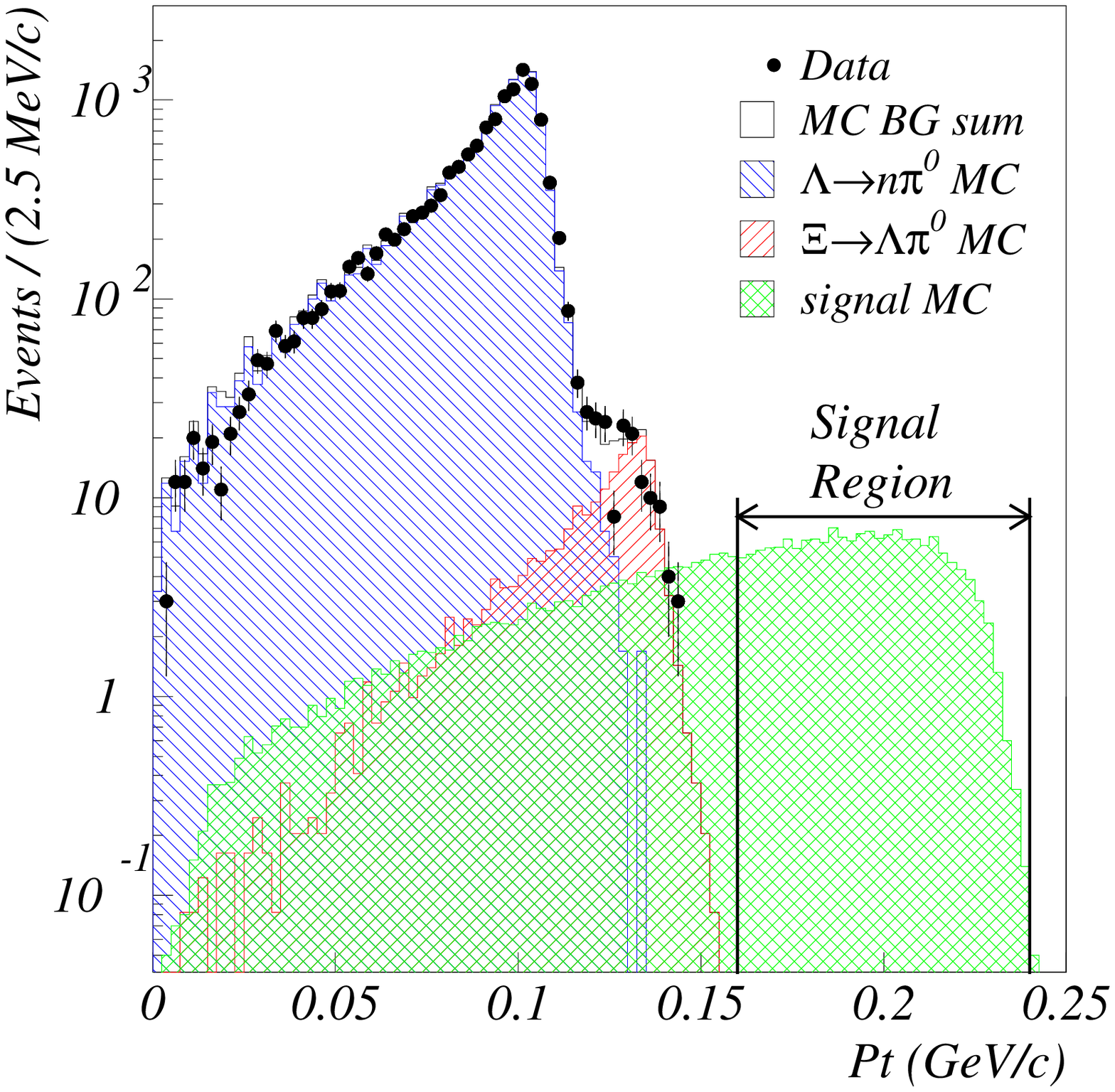,height=1.8in,clip=}}
\begin{center}
\caption{The KTeV apparatus, configured for the rare decay experiment E799, is 
shown on the left. The histogram on the right is the distribution in the 
transverse momentum of the $\mathrm \pi^0$ reconstructed from Dalitz decays, 
with Monte Carlo simulated backgrounds superimposed.}
\label{ktevapp}
\end{center}
\end{figure}

The KTeV apparatus is shown in Figure~\ref{ktevapp}. It consists of a magnetic 
spectrometer and a CsI electromagnetic calorimeter for the detection of charged
decay products and photons, respectively. Particle identification is 
provided by the CsI calorimeter and transition radiation detectors in the case
of electrons and by a hadron filter and muon scintillator hodoscope in the case of 
muons. A set of lead-scintillator
sandwich  $\gamma$ veto counters is situated around the decay region. 
Figure~\ref{ktevapp} shows the distribution in the 
$\mathrm p_T$ of the $\mathrm \pi^0$s detected using Dalitz decays. 
Based on no detected events in the interval 
$\mathrm 160~MeV/c < p_T < 240~MeV/c$, a limit was set~\cite{Alavi-Harati:2000} 
$\mathrm B(K_L^0 \rightarrow \pi^0 \nu \overline{\nu}) < 5.9 \times 10^{-7}$ [90\% CL]. 
A somewhat worse limit, 
$\mathrm B(K_L^0 \rightarrow \pi^0 \nu \overline{\nu}) < 1.8\times 10^{-6}$ [90\% CL] 
was set~\cite{Adams:1999} by the KTeV collaboration using 
$\mathrm 2 \gamma$ decays of the $\mathrm \pi^0$, in which 
one event consistent with background was found.
The limits are well above the SM prediction of $\mathrm \sim 3 \times 10^{-11}$. 

Further significant improvement in sensitivity will come only from dedicated experiments,
and two such experiments are contemplated. E926~\cite{RSVP:1999, E926:1996} at BNL proposes to use a low energy 
pulsed beam and a detector capable of measuring $\gamma$ positions and directions so
as to reconstruct the $\mathrm K_L^0$ decay position and time, and hence determine
the $\mathrm K_L^0$ momentum by time of flight. This allows, in principle, complete determination of
the decay kinematics. Coupled with a hermetic photon veto, the proponents believe
they can detect $\mathrm \sim 65$ events with a signal/background ratio  
greater than two. Significant experimental challenges arise in achieving the 
$\gamma$ veto efficiency needed and in measuring the $\gamma$ positions, angles, and 
times to the required precision. Additionally, the very intense neutral beam
(a mix of kaons, neutrons, and photons) results in 
the potential for large dead-times due to the requirement that events with as 
little as 1-2 MeV energy deposited in an extensive veto system be rejected. This 
experiment is being considered for funding by the U.S. DOE and NSF. 

The KAMI collaboration is studying the possibility of mounting an 
experiment~\cite{KAMI:1997} at Fermilab with approximately the same sensitivity goal.
 It would use
a higher energy beam. The proponents believe the superior $\gamma$ veto efficiency 
that in principle can be achieved at higher energy will allow a large signal to 
background ratio to be achieved without the $\mathrm K_L^0$ momentum determination  
using time of flight, which is not possible at high energy, and without the $\gamma$ 
direction determination, which could be done. Events would be isolated by very 
hermetic $\gamma$ vetos of high efficiency and by inferring the $\mathrm K_L^0$
decay position by containing the invariant mass of two detected $\gamma$'s to 
the $\pi^0$ mass, and then selecting events with a $\pi^0$ transverse momentum
above that from $K_L^0 \rightarrow \pi^0 \pi^0$ decays. This collaboration has
an approved R\&D program and anticipates producing a proposal early in 2001 if 
it is successful. 

\subsection{Search for $\mathbf K_L^0 \rightarrow \pi^0 e^+ e^-$ and 
$\mathbf K_L^0 \rightarrow \pi^0 \mu^+ \mu^-$}

In principle, a measurement of $\mathrm B(K_L^0 \rightarrow \pi^0 l^+ l^-)$
would allow the determination of the CKM parameter $\mathrm \eta$ since there is 
a significant direct CP violating contribution. The rate is about one third that
for  $\mathrm K_L^0 \rightarrow \pi^0 \nu \overline{\nu}$, since there are three 
final states (three neutrino flavors) in the latter. There are also contributions 
from indirect CP violation and from CP conserving amplitudes with $\mathrm 2 \gamma$ 
intermediate
states. The experimental situation is even more difficult due to backgrounds from less
interesting processes, which are discussed below. 

\begin{figure}[htb]
\hbox{
\epsfig{figure=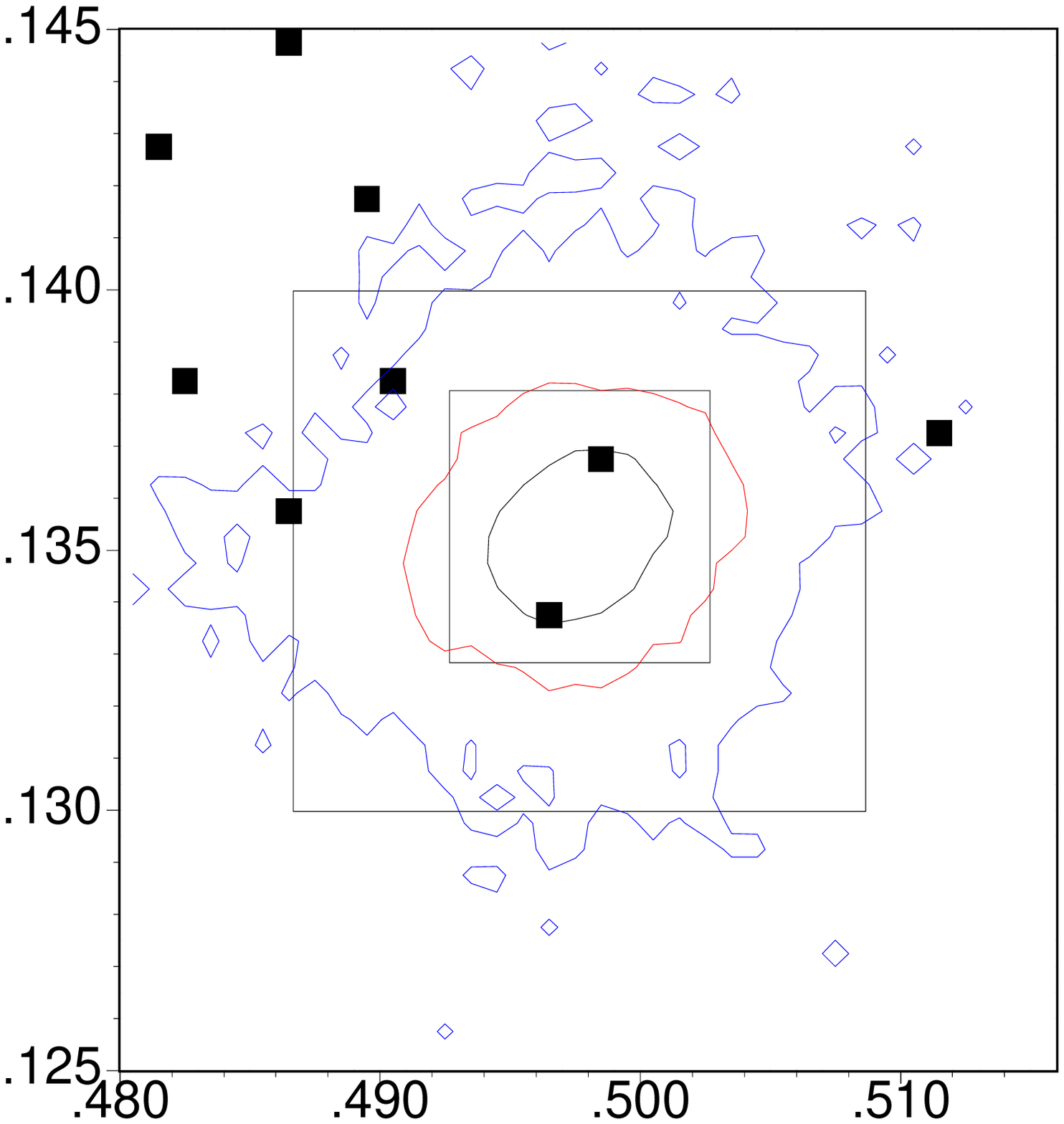,height=1.8in,width=2.5in,clip=} 
\hspace{.2in}
\epsfig{figure=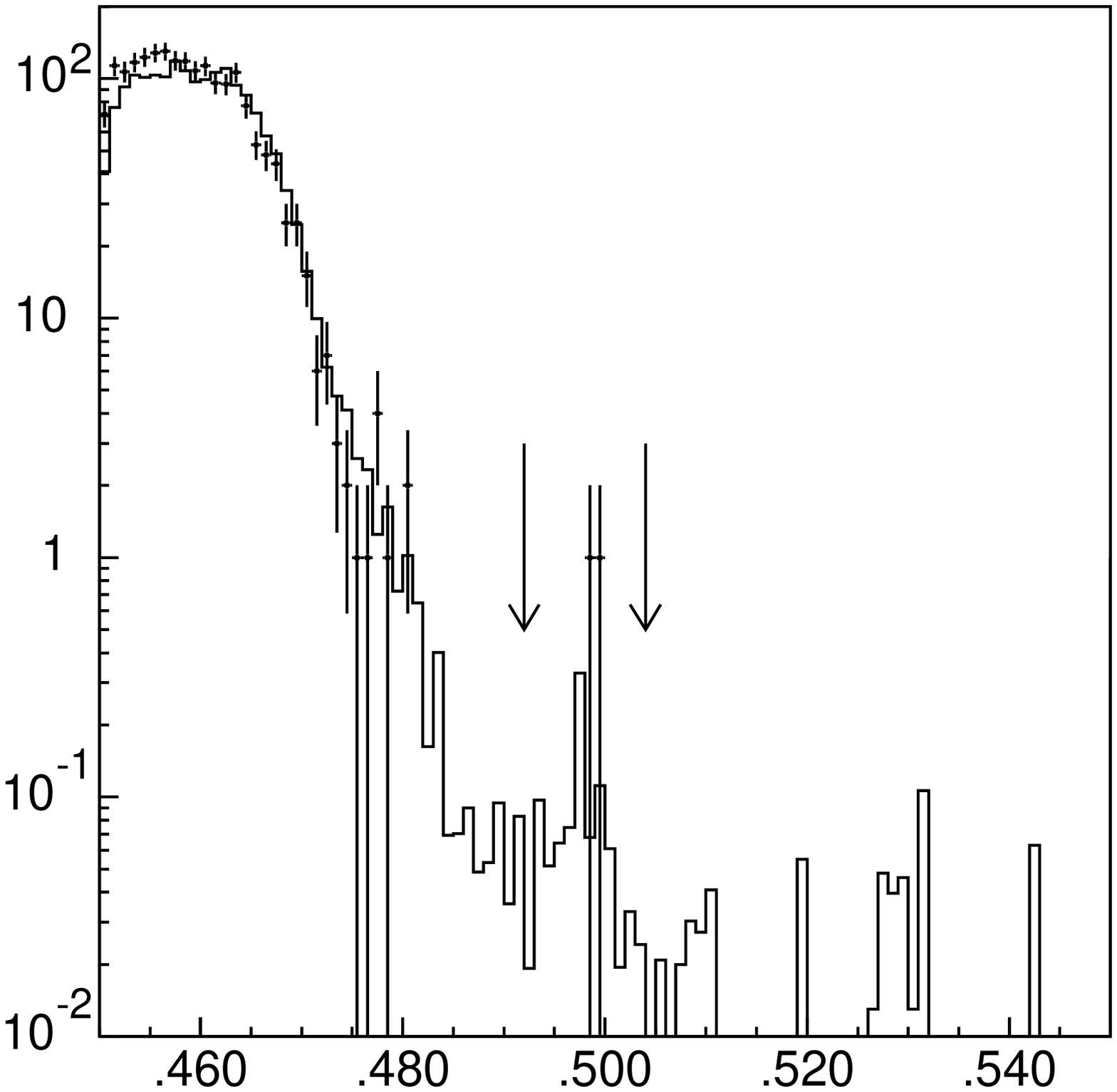,height=1.8in,width=2.5in,clip=}}
\hbox{\hspace{.25in} 
{\small{$\mathrm M_{\gamma \gamma} ~[GeV/c^2]~vs.~M_{ee \gamma \gamma} ~[GeV/c^2]$   
\hspace{1.in}
$\mathrm M_{\mu \mu \gamma \gamma} ~[GeV/c^2]$ }}}
\begin{center}
\caption{On the left is a scatter plot of 
$\mathrm M_{\gamma \gamma} ~vs. ~M_{e e \gamma \gamma}$ for events satisfying all 
selection criteria. The contours correspond to acceptance levels of 68\%, 95\% and 
99\%. On the right is the distribution in $\mathrm M_{\mu \mu \gamma \gamma}$ for 
events satisfying all selection criteria. The final selection on this quantity is 
indicated by the arrows.}
\label{pi0llresults}
\end{center}
\end{figure}

The KTeV collaboration has recently reported results~\cite{Whitmore:1999} 
of a search for both modes.
The apparatus was identical to that used for the 
$\mathrm K_L^0 \rightarrow \pi^0 \nu \overline{\nu}$ search discussed in the 
previous section. Figure~\ref{pi0llresults} shows data for these search modes after
final selection criteria were applied. In both searches, two events are within the 
signal region and consistent with the expected background level. The resulting
limits on the branching fractions are $\mathrm B(K_L^0 \rightarrow \pi^0 e^+ e^-) <
5.64 \times 10^{-10}$ and 
$\mathrm B(K_L^0 \rightarrow \pi^0 \mu^+ \mu^-) < 3.4  \times 10^{-10}$,
both at 90\% CL.   
Both searches are limited by backgrounds. In the case of 
$\mathrm K_L^0 \rightarrow \pi^0 e^+ e^- $, it is dominated by $\mathrm \gamma e^+
e^-$ decay with final state radiation, and is minimized by kinematic cuts to eliminate
events in which one of the photons is either along the direction of the 
$\mathrm e^=$ or $\mathrm e^-$ or in which the a photon momentum is opposite that of 
the $\mathrm e^+ e^-$ 
pair in the $\mathrm K_L^0$ center of mass. The superb resolution of the KTeV 
calorimeter helps in minimizing this background by allowing a relatively small signal 
box; nonetheless, the background is about 100 times the expected signal. In the 
case of  $\mathrm K_L^0 \rightarrow \pi^0 \mu^+ \mu^-$, the background is 
from $\mathrm K_L^0 \rightarrow \pi^0 \pi^+ \pi^-$ decay in which both the 
$\mathrm \pi^+$ and $\mathrm \pi^-$ decay.
In this respect the KTeV apparatus is not well optimized, 
in that $\mathrm \pi$ decay in the center of the analyzing magnet can go undetected 
and result in an apparently large muon momentum and hence large value for the 
reconstructed $\mathrm K_L^0$ mass. Eliminating this type of $\mathrm \pi$ decay 
background was the motivation for two successive magnets in the BNL E871 search for
$\mathrm K_L^0 \rightarrow \mu^{\pm} e^{\mp}$.  
\section{Summary}
There has been significant recent progress in experimental tests of lepton flavor 
conservation using kaons and in measurements of rates for rare decays of kaons from 
which quark mixing matrix elements can be deduced. LFV in the charged sector has been 
searched for in  $\mathrm K_L^0 \rightarrow \mu^{\pm} e^{\mp}$ and 
$\mathrm K^+ \rightarrow \pi^+ \mu^+ e^-$ decays and improved upper limits 
have been set. The expected rate from loop diagrams with neutrino mixing, 
would contribute to this class of LFV processes at a rate well below conceivably 
observable levels. Nearly all models for new physics allow the possibility of LFV, 
and the current upper limits further restrict such models at a mass scale of order 
100 TeV/c$^2$ for full mixing.

The progress in CKM matrix element measurements using rare decays has been in modes 
involving K decays to either lepton pairs 
or a pion and lepton pairs, including the first observation of $\mathrm K_L^0 \rightarrow
e^+ e^-$, a precise measurement of $\mathrm B(K_L^0 \rightarrow \mu^+ \mu^-)$, and
the first observation of $\mathrm K^+ \rightarrow \pi^+ \nu \overline{\nu}$. Improved 
upper limits on $\mathrm K_L^0 \rightarrow \pi^0 {\mathit l \overline{l}}$ have 
also been set. Although the 
current sensitivities for the $\mathrm K \rightarrow \pi {\mathit l \overline{l}}$
modes are insufficient to provide an incisive test of standard model 
predictions, important progress is being made in improving experimental 
techniques and new experiments are proposed that could provide such a test. 

I am grateful to S. Kettell, Y. Wah, J. Whitmore and M. Zeller for providing results 
(in some cases unpublished) of their experiments and information on new proposals.  






\end{document}

%% file: LP99macros.tex
\def\beq{\begin{equation}}
\def\eeq#1{\label{#1}\end{equation}}
\def\eeqn{\end{equation}}

\def\beqa{\begin{eqnarray}}
\def\eeqa#1{\label{#1}\end{eqnarray}}
\def\eeqan{\end{eqnarray}}

\let\bar=\overbar

\def\Dslash{\not{\hbox{\kern-4pt $D$}}}
\def\dslash{\not{\hbox{\kern-2pt $\del$}}}

\def\msb{{\bar{\ssstyle M \kern -1pt S}}}

